\documentclass[11pt]{article}
\usepackage{doc}
\usepackage{makeidx}

\usepackage{epsfig}
\usepackage{subfigure} 
\usepackage{graphicx} 
\usepackage{dcolumn}
\usepackage{bm}

\begin{document}

\newcommand\grad{\nabla} \newcommand\f{\frac} \newcommand\old[1]{{ }}
\newcommand\divg{\nabla\cdot} \newcommand\curl{\nabla\times}
\newcommand\p{\partial} \newcommand\B{\mathbf{B}}
\newcommand\E{\mathbf{E}} \newcommand\J{\mathbf{J}}
\newcommand\vel{\mathbf{u}} \newcommand\normal{\mathbf{n}}
\renewcommand\H{\mathbf{H}} \newcommand\be{\begin{equation}}
\newcommand\en{\end{equation}}

\newcommand\hf[1]{\frac{#1}{2}} \newcommand\inv[1]{\frac{1}{#1}}
\newcommand\half{\frac{1}{2}}

\newcommand\diff[2]{\frac{\textrm{d}{#1}}{\textrm{d}{#2}}}
\newcommand\diffsec[2]{\frac{\textrm{d}^2{#1}}{\textrm{d}{#2}^2}}
\newcommand\diffp[2]{\frac{\partial{#1}}{\partial{#2}}}
\newcommand\diffpa[3]{\left.\diffp{#1}{#2}\right|_{#3}}

\title{\bf Transonic ablation flow regimes of high-Z pellets}

\author{ \bf Hyoungkeun Kim$^{1}$, Roman Samulyak$^{1,2}$, Paul
  Parks$^{3}$ \\ \it $^{1}$Department of Applied Mathematics and
  Statistics, \\ \it Stony Brook University, Stony Brook, NY 11794
  \\ \it $^{2}$Computational Science Center, \\ \it Brookhaven
  National Laboratory, Upton, NY 11973 \\ \it $^{3}$General Atomics,
  P.O. Box 85608, \\ \it San Diego, California 92186-5608}

\date{July 15, 2014}%
\maketitle

The injection of cryogenic pellets for plasma refueling is the method
of choice for future reactor-scale tokamaks such as ITER
\cite{Pegourie}. One-dimensional, spherically-symmetric, quasi-steady
state models such the transonic flow model \cite{Parks78} have
provided scaling laws and successfully predicted pellet ablation rates
and penetration depths. Theoretical work has been complemented by
two-dimensional numerical simulations that resolved detailed physics
processes in the ablation cloud \cite{Ishizaki04}, including the MHD
effects \cite{SamLuParks07, ParksLuSam09}.  Previous studies, focusing
on the fueling application of pellets, have limited the choice of the
pellet material to deuterium.  However high-Z pellets have a great
potential for other important tokamak applications, in particular the
plasma disruption mitigation.  In this letter, we report results of
numerical studies of the ablation of argon and neon pellets and
compare them with theoretical predictions and studies of deuterium
pellets. Simulations have been performed in the spherically-symmetric
approximation using the hydrodynamic / MHD code FronTier
\cite{FT_code} with recently developed physics models for the pellet
ablation such as the electronic heat flux model and the numerical
equation-of-state (EOS) with the support for multiple ionization of
high-Z gases \cite{KimSamZhang12}. The code has been extencively used
for simulations of phase transitions \cite{RS7}, high power mercury
target experiments \cite{RS14,RS15}. An overview of application
problems is given in \cite{TSTT_code}.

The main equations for the pellet ablation cloud are the inviscid
Euler equations with external heat source:
\begin{eqnarray}
\frac {\p \rho}{\p t} & = & -
\nabla\cdot(\rho\vel),\label{eq_rho}\\ \rho \left(\frac \p{\p t} +\vel
\cdot\nabla\right)\vel & = & - \nabla P,\label{eq_m}\\ \rho
\left(\frac \p{\p t} +\vel \cdot\nabla\right)e & = & -P\nabla\cdot\vel
- \grad \cdot {\bf q},\label{eq_e}\\ P & = &
P(\rho,e)\label{eq_closer}
\end{eqnarray}
where $\rho$, $\vel$, $e$, and $P$ are density, velocity, specific
internal energy, and pressure, respectively. The term
($-\grad\cdot{\bf q}$) described the external heat source of hot
electrons streaming along magnetic field line into the ablation cloud.

\begin{figure}\begin{center}
\subfigure[Average ionization]{\label{m_GA}\psfig{figure=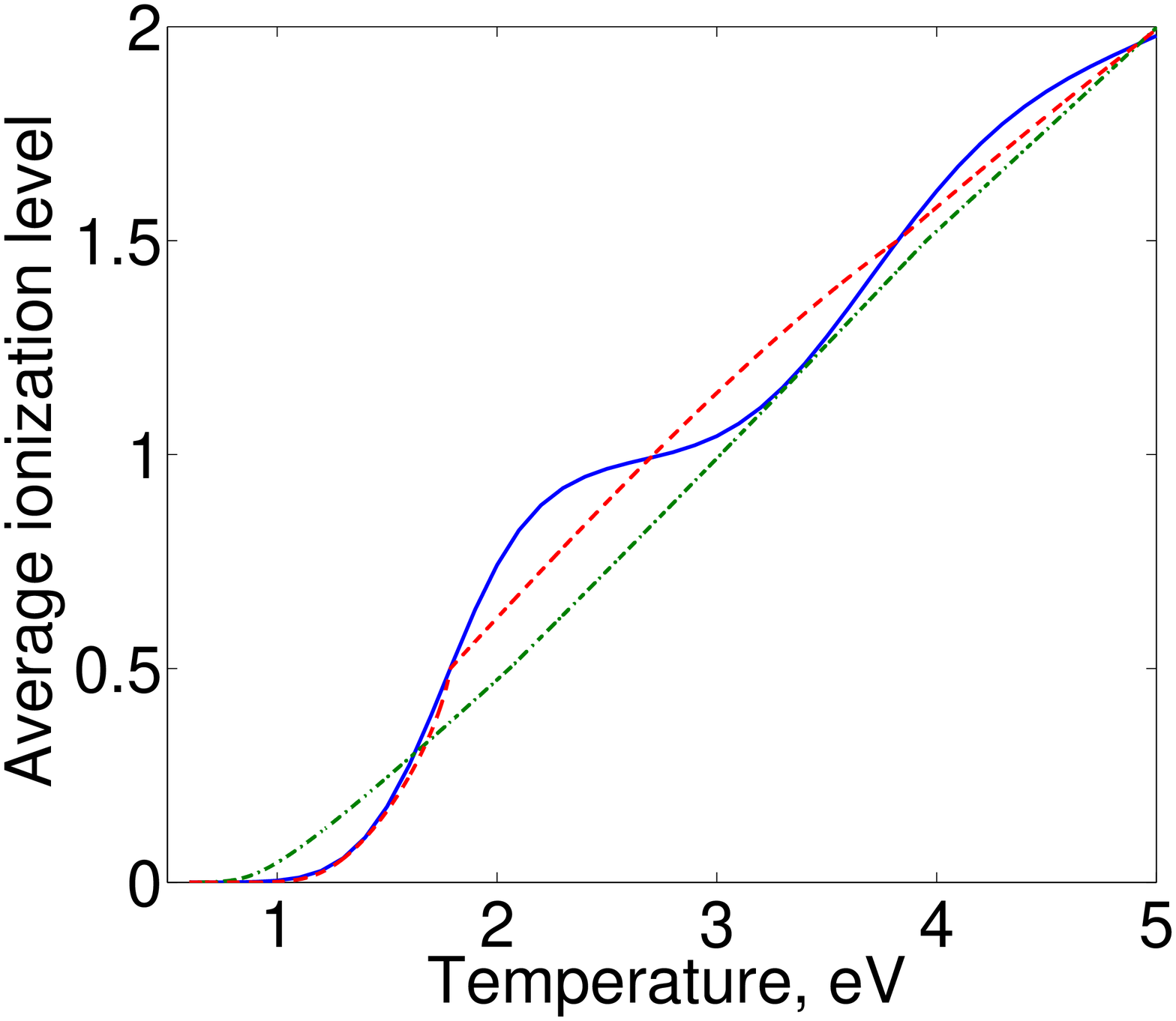,width=0.475\textwidth}}
\subfigure[Specific internal energy]{\label{e_GA}\psfig{figure=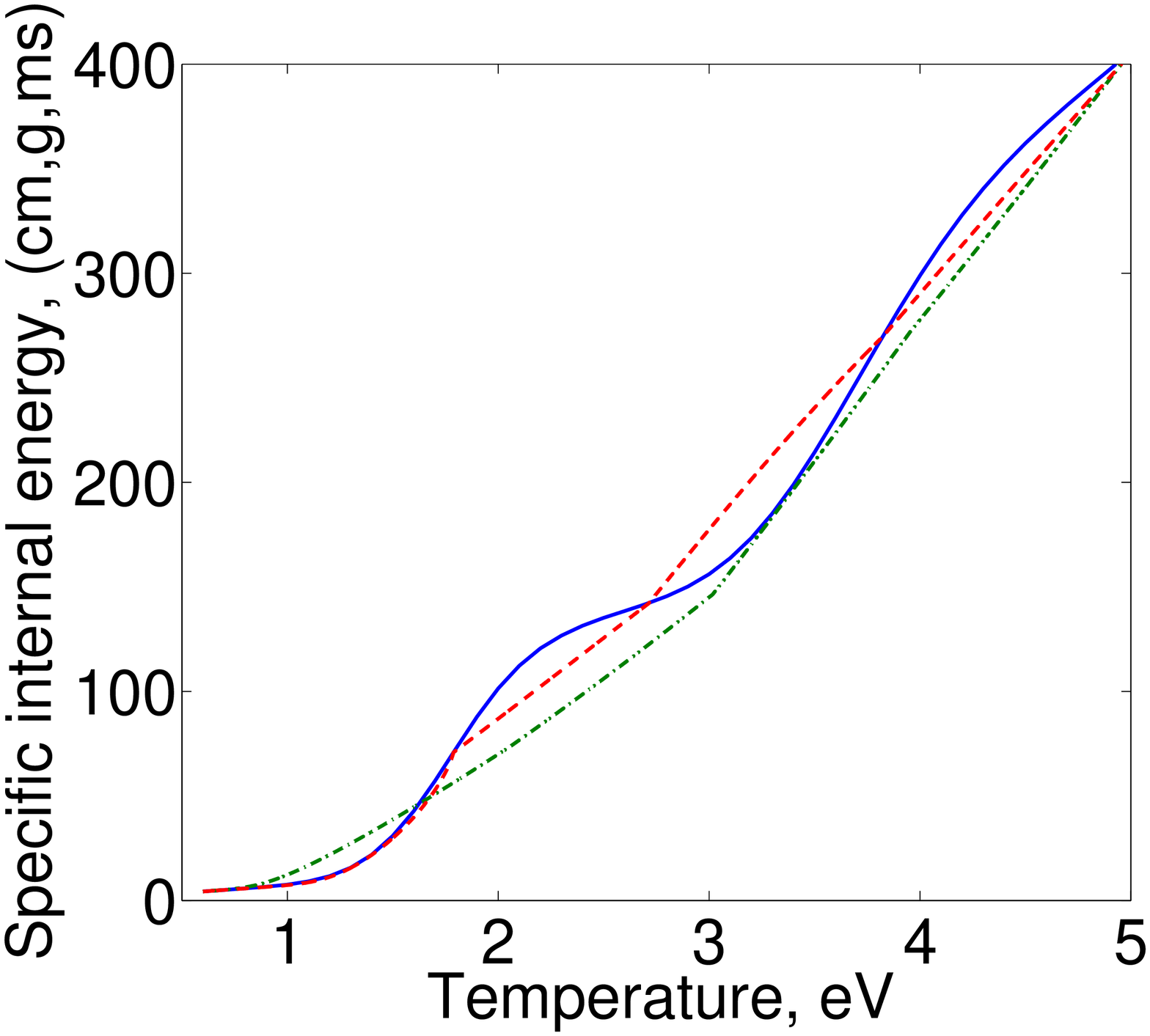,width=0.475\textwidth}}
\caption{The comparison of solutions from the coupled system of Saha
  equations (solid blue line), the original average ionization model
  (green dash-dotted line), and the improved average ionization model
  (red dashed line) for neon with density of
  $3.351\times10^{-5}~g/cm^3$.}
\label{m_e_comparison_GA}
\end{center}
\end{figure}

\begin{figure}\begin{center}
\subfigure[Ionization energy,
  $\overline{I}$]{\label{I_m}\psfig{figure=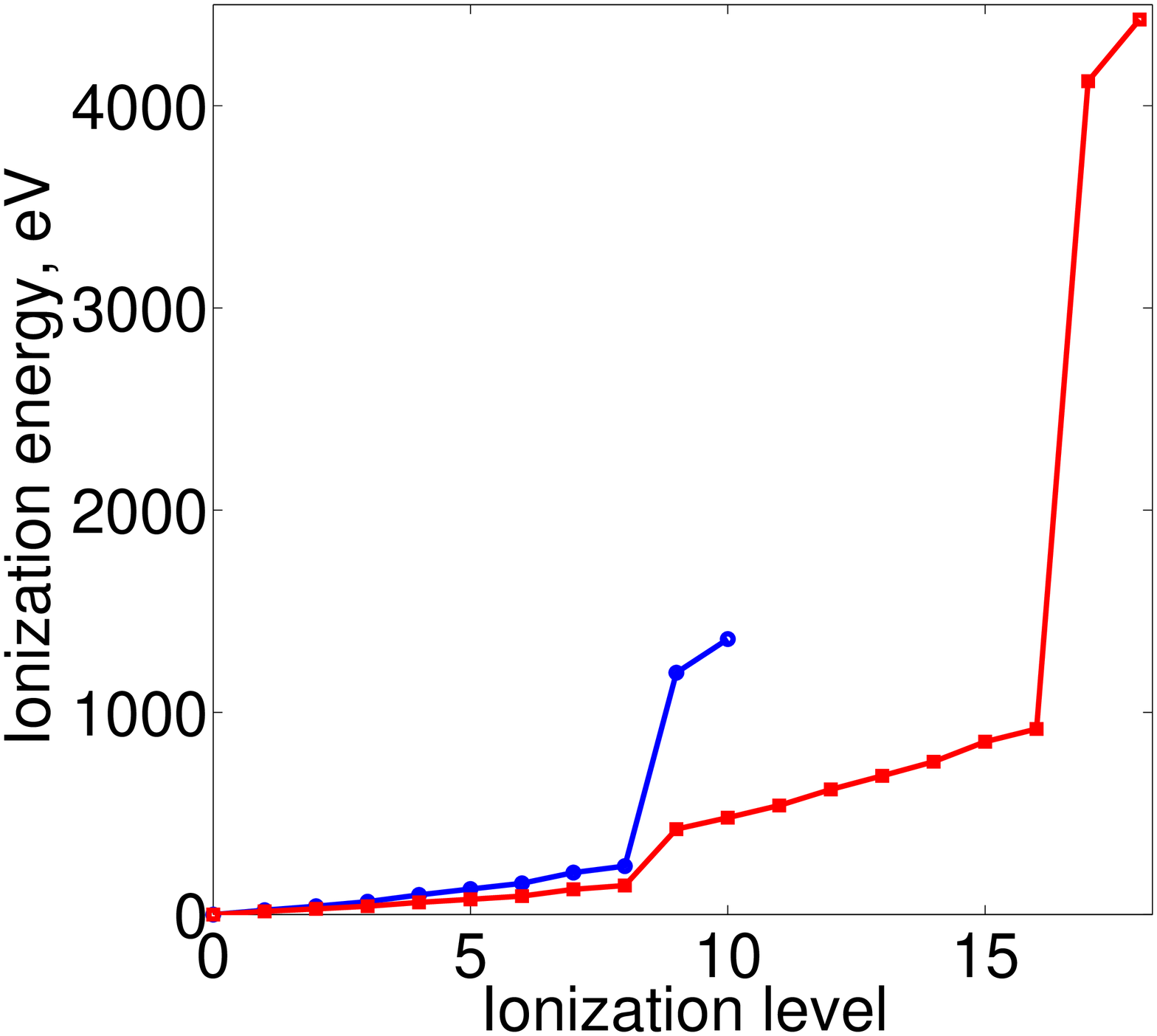,width=0.475\textwidth}}
\subfigure[Statistical weight ratios,
  $\overline{u}$]{\label{SW_m}\psfig{figure=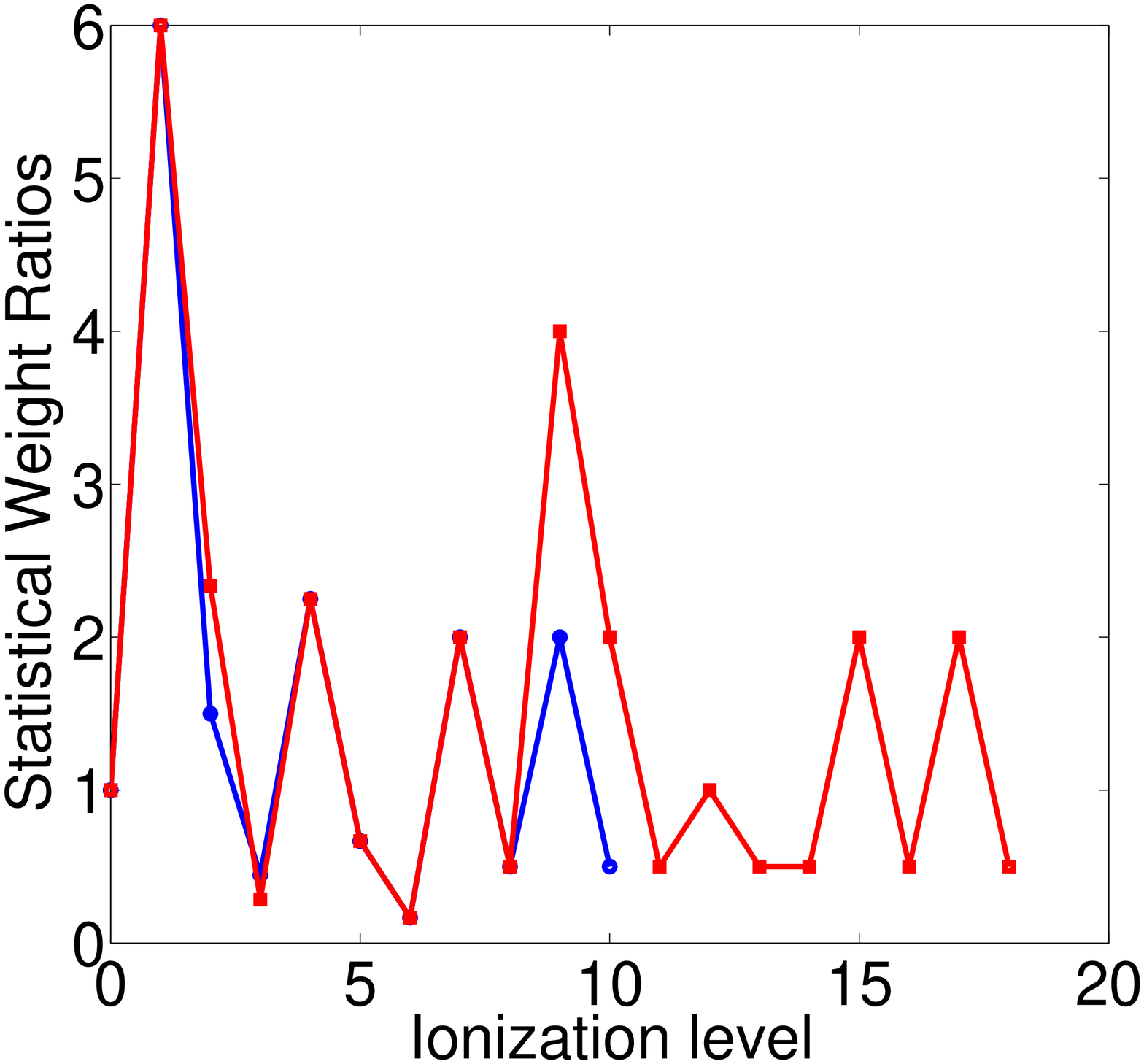,width=0.475\textwidth}}
\caption{Ionization energy and statistical weight ratios of neon
  (blue-circle) and argon (red-square)}
\label{I_SW_m_Ne_AR}
\end{center}
\end{figure}

As details of transonic regimes of the pellet ablation flow are
strongly dependent on atomic physics processes in the ablation cloud,
the quality of numerical equation of state models describing partially
ionized plasmas is of significant importance.  While probabilities of
multiply ionized states in high-Z materials in the local thermodynamic
equilibrium is accurately described by coupled system of Saha
equations \cite{Zeldovich}, the direct use of these equations in
time-dependent hydrodynamic simulations is prohibitively
computationally intensive. In our previous works \cite{KimSamZhang12,
  KimZhangSam13}, we have developed a numerical EOS model for argon
based on the Zel$'$dovich approximation of average ionization
\cite{Zeldovich}. The average ionization model (AIM) relates the
average ionization level $\overline{m}$ with the thermodynamic states
as $\overline{m} = \f{AT^{3/2}}{n}\exp{-\f{\overline{I}}{kT}}$ and
computes the pressure as $P=(1+\overline{m})\rho RT$.  This model is
sufficiently accurate in the high energy regime. However, as it is
shown in Figure \ref{m_e_comparison_GA}, the accuracy is reduced at
low temperatures corresponding to the average ionization level of
order ($\overline{m}\sim0.1$).  The ablation cloud near the pellet
surface is in weakly ionized state, and the inaccuracy of the original
AIM could change the overall dynamics of ablation cloud in
simulations. In this work, we improve our previous numerical EOS for
high-Z gases by including the linear continuum approximation of
statistical weight ratio ($\overline{u}$) shown in Figure
\ref{I_SW_m_Ne_AR}. The equation in the improved AIM with
$\overline{u}$ is: \be \overline{m} =
\overline{u}~\f{AT^{3/2}}{n}\exp{-\f{\overline{I}}{kT}}. \label{bar_m}
\en The accuracy of this numerical EOS is increased further by using
the numerical optimization of $\overline{I}$ when $\overline{m} <
1$. We use the modified $\overline{I}$ instead of the linear
approximation to mimic the single electron model in the low energy
regime. In this work, $\overline{I}=I_1(1-\overline{m}^{q})^{1/q}$ is
used, where $I_1$ is the first ionization energy and $q$ is a properly
picked rational number when $\overline{m} < 1$. The result of the
improved AIM is shown in the Figure \ref{m_e_comparison_GA}.

We compare simulations of high-Z gases with molecular deuterium
described in \cite{Ishizaki04,SamLuParks07,ParksLuSam09}.  The
equation of state model for deuterium that accounts for dissociation
and ionization uses the exact system of Saha equations.  We also
compare our results with simulations obtained using the polytropic EOS
model that defines the gas pressure as $P=(\gamma-1)\rho e$, where
$\gamma$ is the ratio of specific heats.
  
The electron heat flux model is identical to that of the previous
works \cite{ParksSesBay00, Ishizaki04, SamLuParks07, ParksLuSam09}
except the modification of the Coulomb logarithm and the dimensionless
opacity for high-Z atoms. The modified Coulomb logarithm is:
 \[
\ln\Lambda = \frac{\overline{m}}{Z}\ln\Lambda_{ef} +
\frac{(1-\overline{m})}{Z}\ln\Lambda_{eb}
\]
in which $Z$=atomic number, $\Lambda_{ef}=0.2E/\hbar\omega_{pe}$,
$\omega_{pe}=(4\pi n_ee^2/m_e)^{1/2}$, $\Lambda_{eb}=E/I_*\sqrt{e/2}$,
$E\approx2T_{e\infty}$(plasma electron temperature), $n_e$=plasma
electron density, and $I_*$ is the mean excitation energy (for
example, $I_*^H = 19.2$~eV, $I_*^{Ne} = 137$~eV, and $I_*^{Ar} =
188$~eV from \cite{ICRU84}). The changed dimensionless opacity in the
spherically symmetric approximation is $u = \tau/\tau_{eff}$, where
\[
\begin{array}{l}
\tau(r) = Z\int_{r}^{\infty} n(r')dr', ~~~~~\tau_{eff} =
\tau_{\infty}\sqrt{\frac{2}{1+Z}}, ~~~~~\tau_{\infty} =
\frac{T_{e\infty}^2}{8\pi e^4\ln\Lambda}.
\end{array}
\]
The pellet surface ablation model is identical to that of
\cite{SamLuParks07} with the exception of some technical improvements,
namely the numerical treatment of high gradients of physics quantities
near the pellet surface.

\begin{figure}\begin{center}
\subfigure[Polytropic
  EOS]{\label{D_poly}\psfig{figure=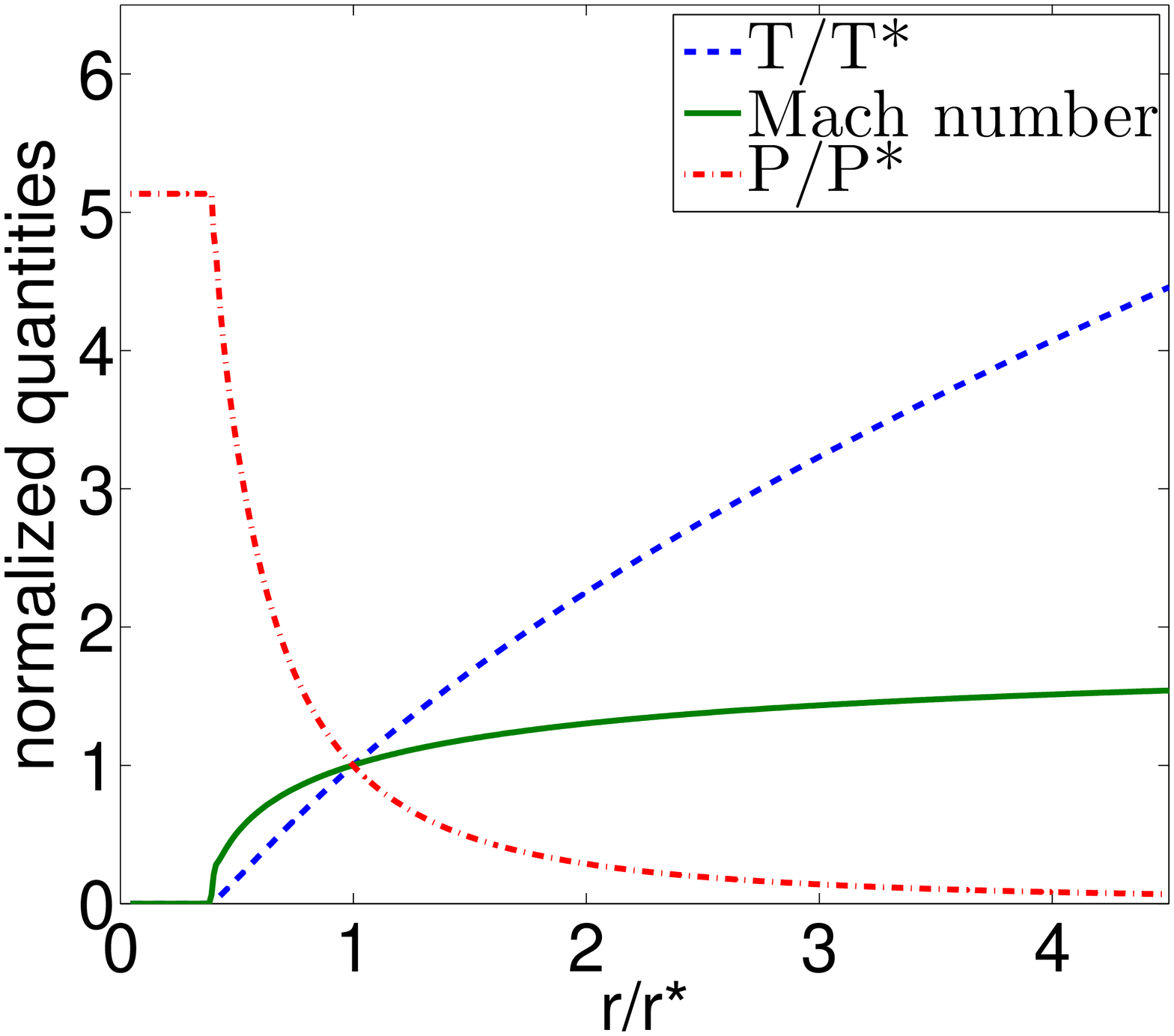,width=0.475\textwidth}}
\subfigure[Plasma
  EOS]{\label{D_plasma}\psfig{figure=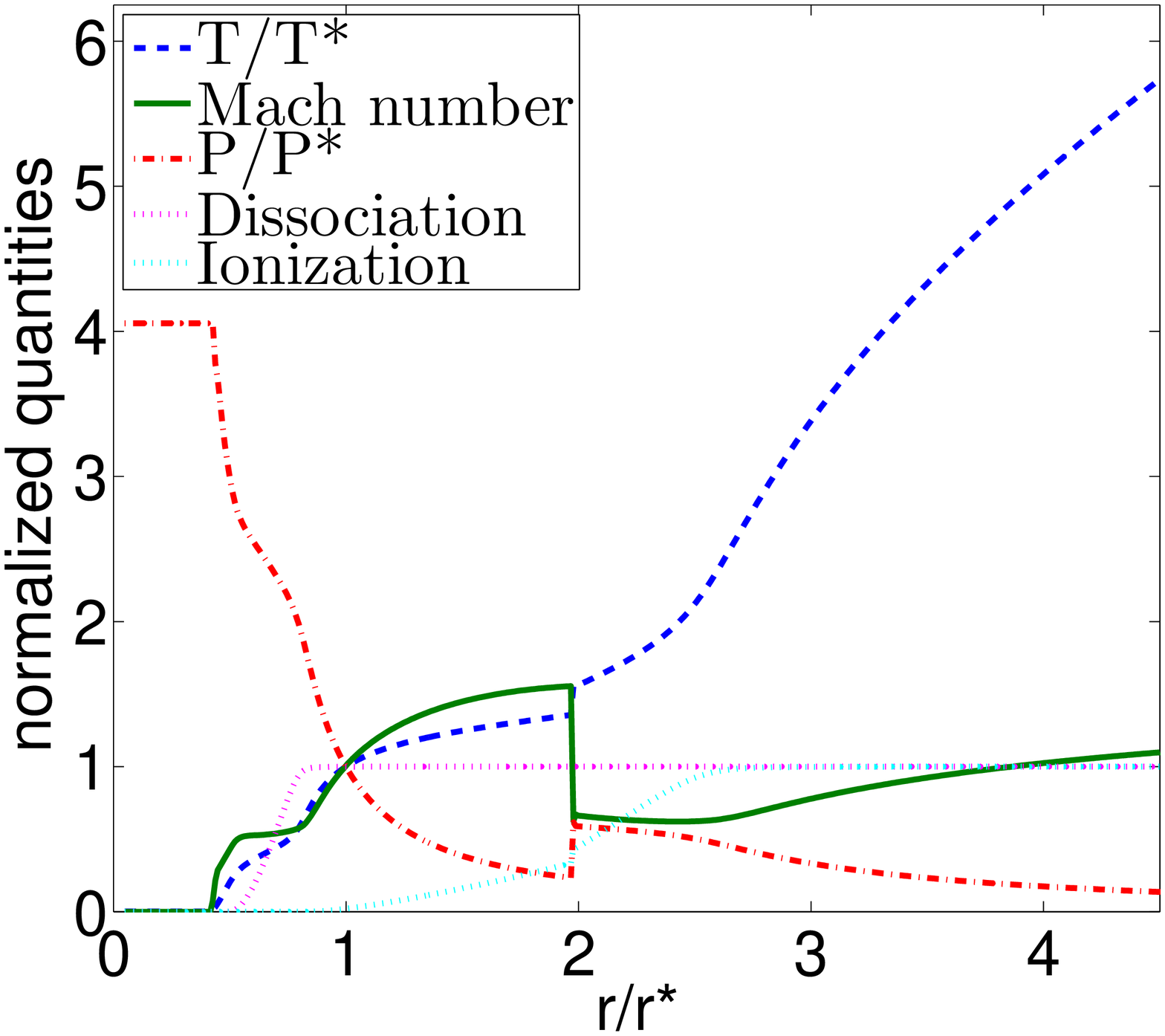,width=0.475\textwidth}}
\caption{Normalized ablated cloud profiles of deuterium pellet in
  1-dimensional spherically symmetric model of ablation (a) without
  atomic processes (polytropic EOS), and (b) with atomic processes
  (plasma EOS).}
\label{D_plot}
\end{center}
\end{figure}

We start with the simulation of a deuterium pellet using parameters of
\cite{SamLuParks07}, namely the pellet radius of $r_p=0.2cm$, the
plasma electron temperature of $T_{e\infty}=2keV$, and the plasma
electron density of $n_{e\infty}=10^{14}cm^{-3}$. Simulations with the
polytropic EOS demonstrate a transonic ablation flow that starts as
subsonic near the pellet surface and changes to supersonic due to the
electron heat flux (Figure \ref{D_poly}).  When the deuterium plasma
EOS is used (Figure \ref{D_plasma}), the ablation flow is affected by
energy sinks due to the dissociation and ionization.  The dissociation
processes slow down the increase of the Mach number near the pellet
surface.  The flow accelerates then to supersonic velocities before
the majority ionization processes occur. The ionization energy causes
the shock wave and the drop of the Mach number below unity. The flow
then accelerates again and reached the supersonic state the second
time (double transonic pellet ablation regime). The temperature $T^*$
and pressure $P^*$ of the ablation cloud at the sonic radius $r^*$ are
shown in Table \ref{table_D}.  The ablation rate using polytropic EOS
in Table \ref{table_D} is consistent with the value of 133 g/s
predicted by the theoretical transonic flow model of \cite{Parks78}.
The ablation rate is reduced by approximately $10.7\%$ when the atomic
processes are included in the EOS.

\begin{table}[!h]
\begin{center}
\begin{tabular}{|c|c|c|c|c|c|}
\hline & r* (cm) & T* (eV) & P* (bar) & Ablation rate (g/s) \\ \hline
Polytropic EOS ($\gamma=7/5$) & 0.518 & 3.21 & 29.25 & 132.6 \\ Plasma
EOS & 0.474 & 1.04 & 25.54 & 118.4 \\ \hline
\end{tabular}
\end{center}
\caption{The ablated cloud states of deuterium pellet at the first
  sonic radius (r*) for the cases with polytropic EOS and plasma EOS.}
\label{table_D}
\end{table}

\begin{table}[!h]
\begin{center}
\begin{tabular}{|c|c|c|c|c|c|}
\hline & r* (cm) & T* (eV) & P* (bar) & $\overline{m}$* & Ablation
rate (g/s)\\ \hline Polytropic EOS ($\gamma=5/3$) & 0.603 & 30.15 &
22.96 & - & 112.9 \\ Plasma EOS & 0.603 & 4.84 & 16.37 & 1.99 & 95.3
\\ \hline
\end{tabular}
\end{center}
\caption{The ablated cloud states of neon pellet at the first sonic
  radius (r*) for the cases with polytropic EOS and plasma EOS.}
\label{table_Ne}
\end{table}

\begin{figure}\begin{center}
\subfigure[Polytropic
  EOS]{\label{Ne_poly}\psfig{figure=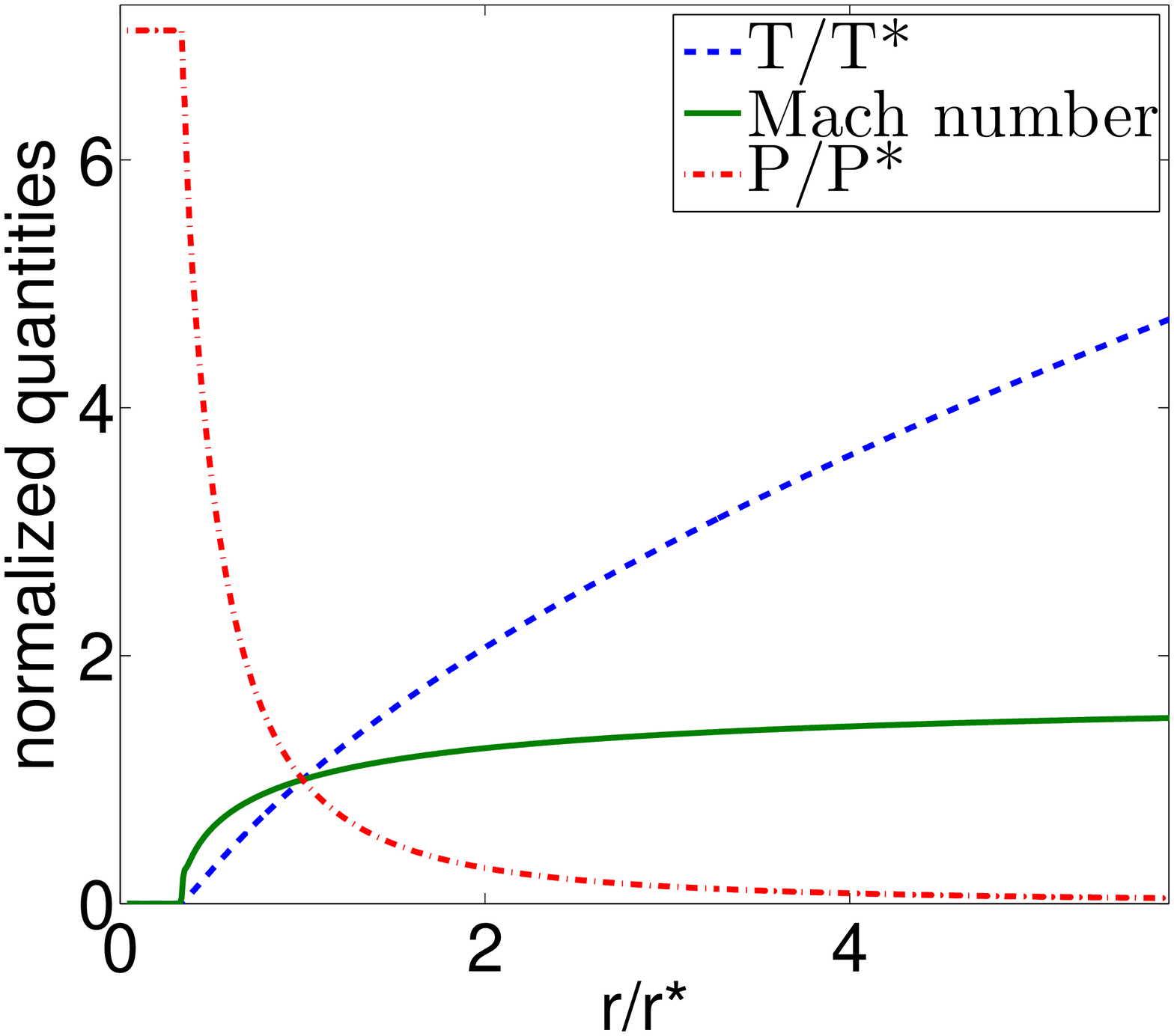,width=0.475\textwidth}}
\subfigure[Plasma
  EOS]{\label{Ne_plasma}\psfig{figure=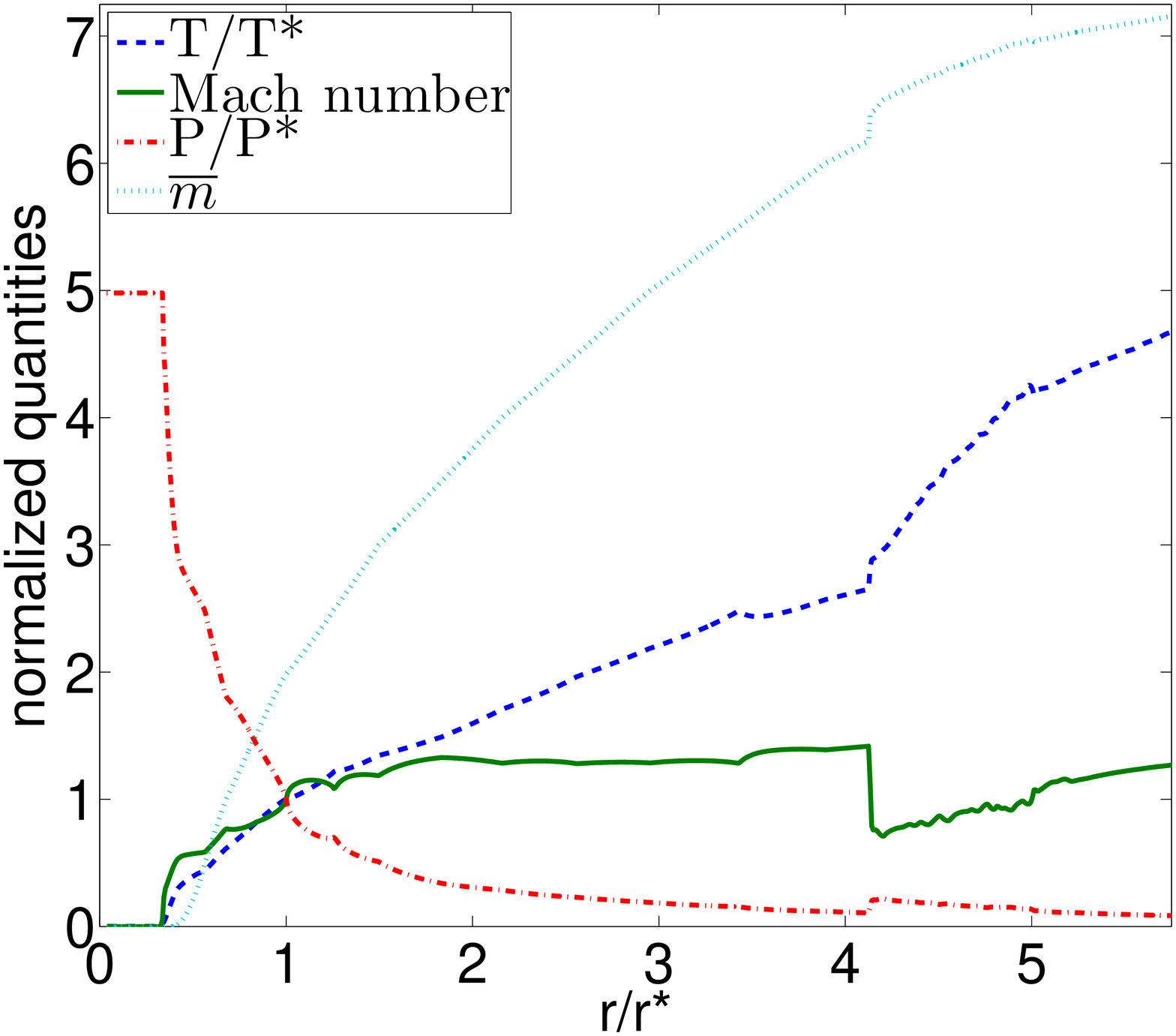,width=0.475\textwidth}}
\caption{Normalized ablated cloud profiles of neon pellet in
  1-dimensional spherically symmetric model of ablation (a) without
  atomic processes (polytropic EOS), and (b) with atomic processes
  (plasma EOS).}
\label{Ne_plot}
\end{center}
\end{figure}

\begin{table}[!h]
\begin{center}
\begin{tabular}{|c|c|c|c|c|c|}
\hline & r* (cm) & T* (eV) & P* (bar) & $\overline{m}$* & Ablation
rate (g/s)\\ \hline Polytropic EOS ($\gamma=5/3$) & 0.588 & 61.82 &
22.62 & - & 103.6 \\ Plasma EOS & 0.415 & 4.46 & 20.04 & 2.49 & 77.4
\\ \hline
\end{tabular}
\end{center}
\caption{The ablated cloud states of argon pellet at the first sonic
  radius (r*) for the cases with polytropic EOS and plasma EOS.}
\label{table_Ar}
\end{table}

\begin{figure}\begin{center}
\subfigure[Polytropic
  EOS]{\label{Ar_poly}\psfig{figure=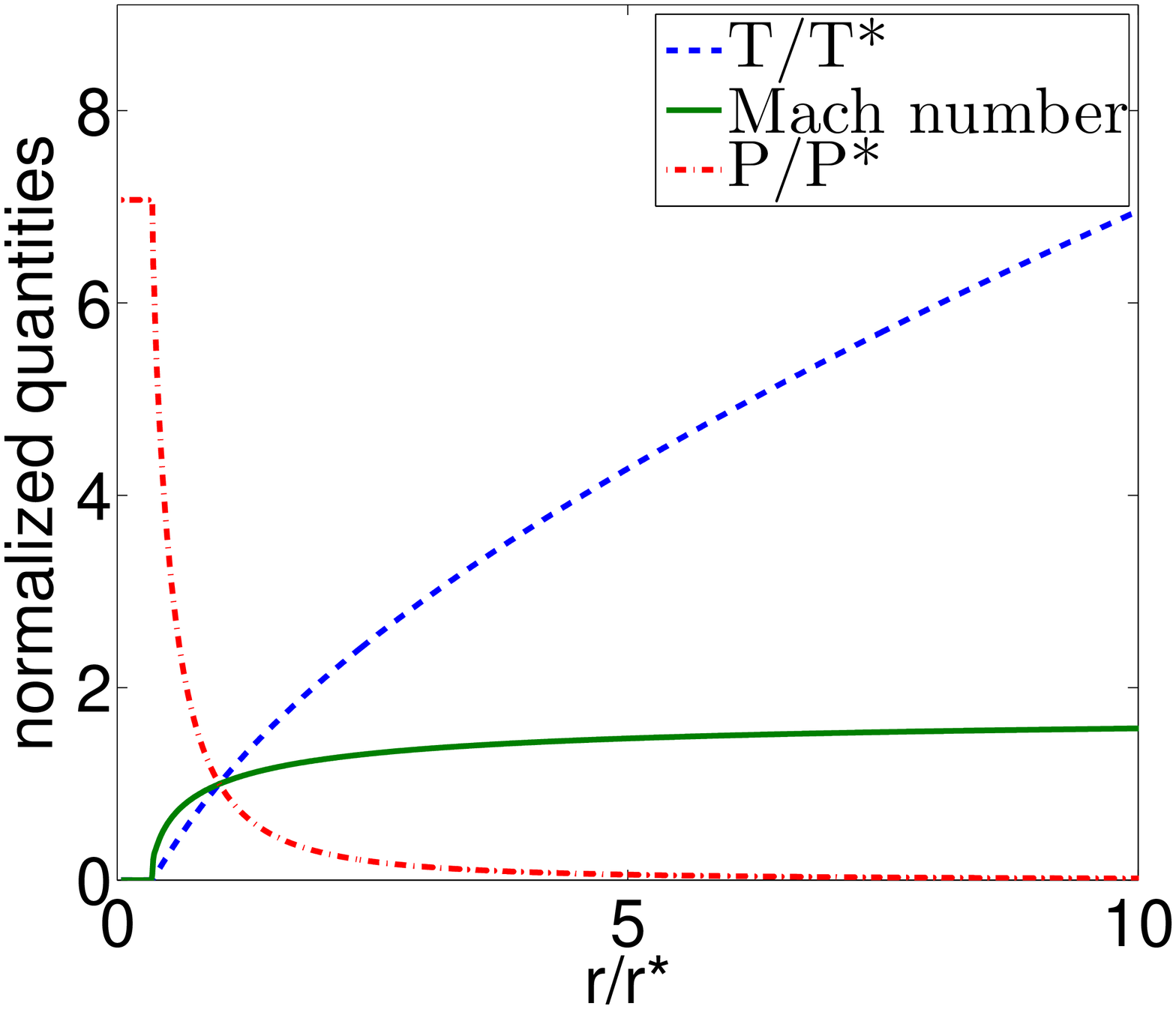,width=0.475\textwidth}}
\subfigure[Plasma
  EOS]{\label{Ar_plasma}\psfig{figure=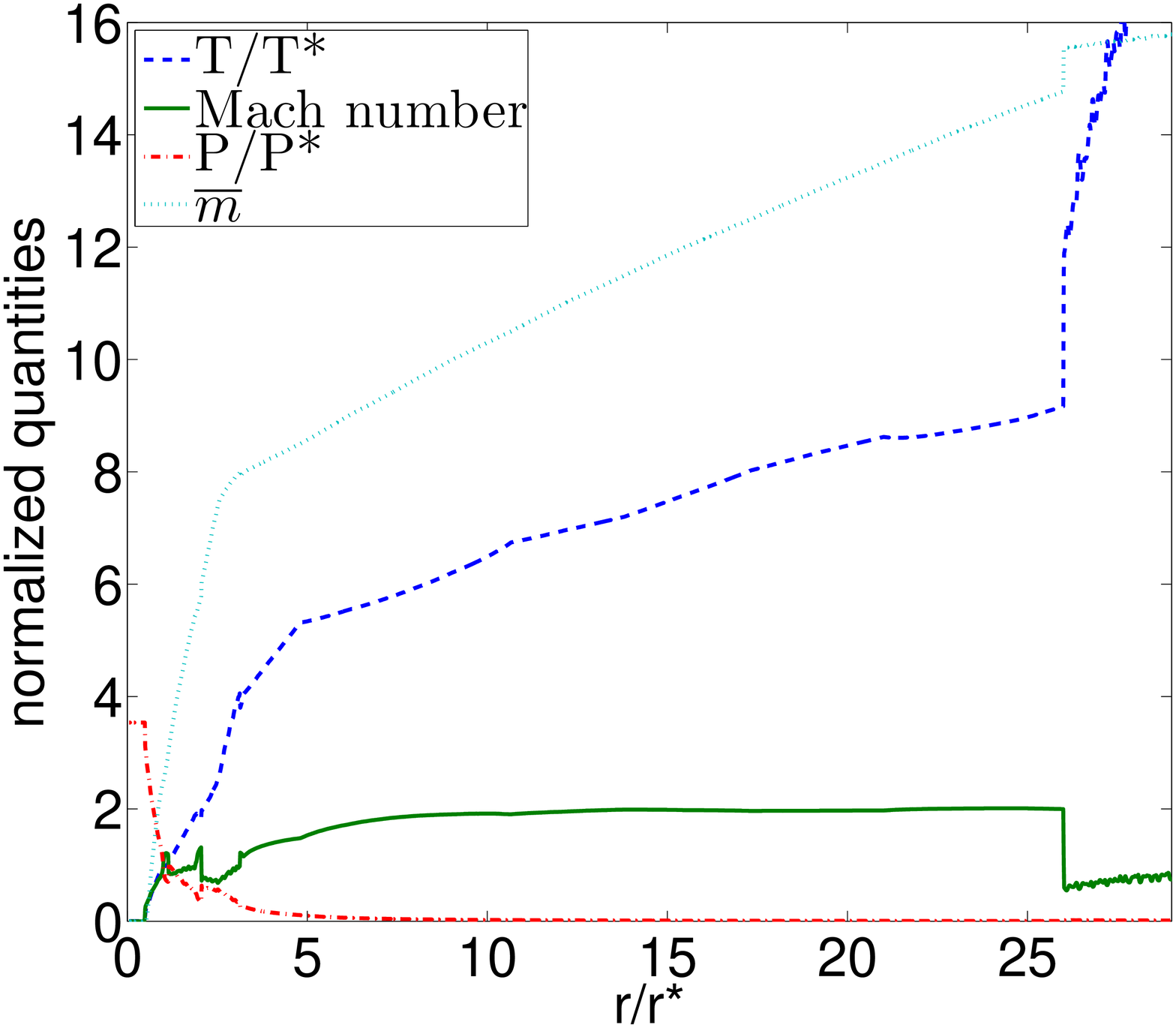,width=0.475\textwidth}}
\caption{Normalized ablated cloud profiles of argon pellet in
  1-dimensional spherically symmetric model of ablation (a) without
  atomic processes (polytropic EOS), and (b) with atomic processes
  (plasma EOS).}
\label{Ar_plot}
\end{center}
\end{figure}

\begin{figure}\begin{center}
\subfigure[Ablation flow of argon
  pellet]{\label{Ar_local}\psfig{figure=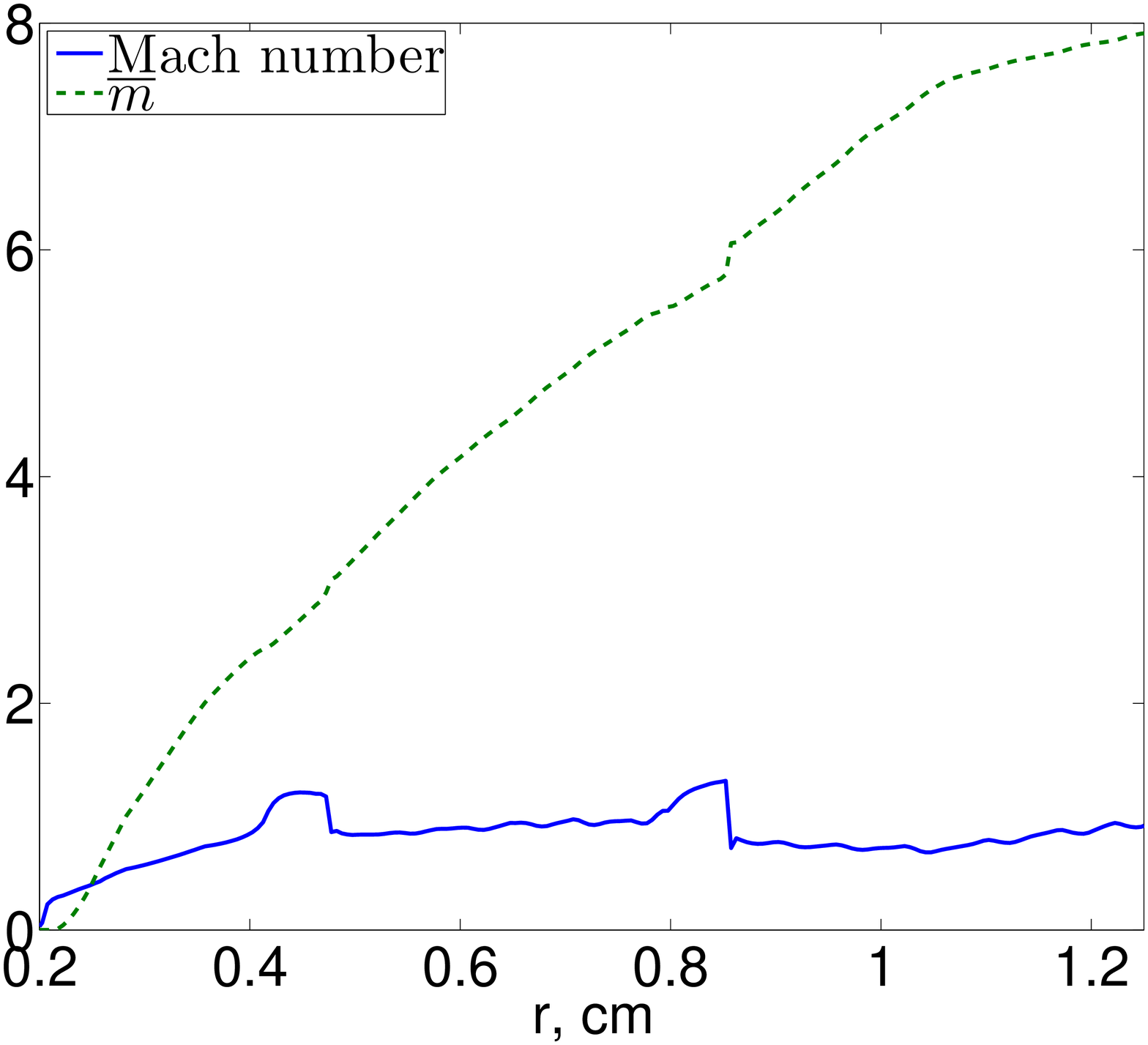,width=0.475\textwidth}}
\subfigure[Atomic property of
  argon]{\label{Ar_I_SWR}\psfig{figure=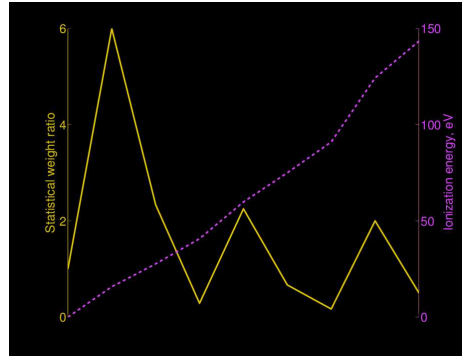,width=0.475\textwidth}}
\caption{The local property of argon pellet's ablation flow near the
  pellet surface (a) and the corresponding atomic property of argon
  (b).}
\label{Ar_local_analysis}
\end{center}
\end{figure}

\begin{figure}\begin{center}
\psfig{figure=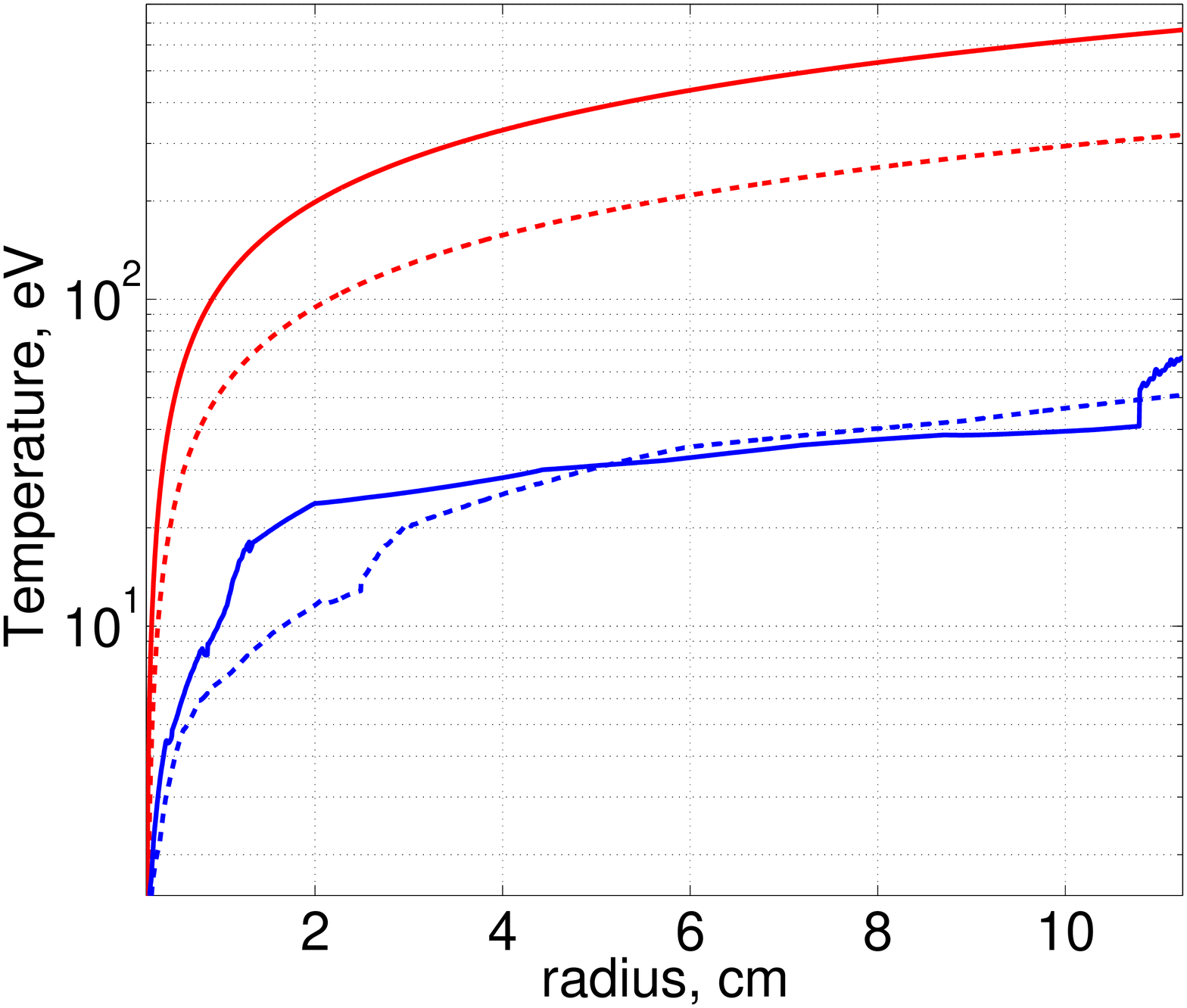,width=0.475\textwidth}
\caption{The comparison of ablated gas temperature of neon pellet
  (dashed line) and argon pellet (solid line) with polytropic EOS (red
  line) and plasma EOS (blue line)}
\label{Ablated_gas_T_compare}
\end{center}
\end{figure}

In the pellet ablation simulation of neon and argon, the states of
ablation cloud at sonic radius is shown in the table \ref{table_Ne}
and \ref{table_Ar} and the normalized quantities based on them are
provided in Figures \ref{Ne_plot} and \ref{Ar_plot}. For simulations
that neglect ionization using the polytropic EOS, the ablation rates
in Tables \ref{table_Ne} and \ref{table_Ar} are in reasonably good
comparison to the theoretical predictions of 109 g/s for neon and 103
g/s for argon using the transonic flow model \cite{Parks78}.  The
ablation flow of the neon pellet exhibits the double transonic regime
similar to the deuterium case.  Slowly increasing ionization
potentials of the neon atom corresponding to the ionization levels
from 1 to 8 irregularly reduce the ablation flow acceleration and
cause oscillations on the Mach number plot in Figure \ref{Ne_plasma}.
Despite these energy sinks, the flow accelerated above the sonic
point. Then the large increase of the ionization energy associated
with stripping off the 9th and 10th electrons drop the Mach number
below unity and cause the shock wave. Then the flow again reaches the
supersonic state.

The ablation flow regime is more complex for the argon pellets (Figure
\ref{Ar_plasma}).  The ionization energy of the argon atom slowly
increases with the increase of the ionization level from 1 to 16, with
the exception of a bigger change of the ionization potential between
levels 8 and 9 (see Figure \ref{I_m}). Nevertheless, the Mach number
twice drops below unity before the average ionization reaches level 6
(see Figure \ref{Ar_local_analysis} that shows details of the Mach
number and the average ionization close to the pellet surface). This
is caused by the combination of the increasing ionization energy with
statistical weights that rapidly increase for the ionization level 3
and 6 (see Figure \ref{Ar_I_SWR}), causing the weak shock waves in the
ablation flow. Then the flow slowly and steadily accelerates under
constant energy removal by ionization. When the ionization level
reaches 16, the rapid increase of the ionization potential between
levels 16 and 17 causes the third shock wave.  Then the flow of fully
ionized argon accelerates again above the sonic point.

A large reduction of temperature in the neon and argon pellet
simulations with real gas EOS compared to the polytropic gas
simulations can be understood via consecutive ionization energy
losses. For the argon pellet, there is one order of magnitude
difference of temperature at the first sonic radius (table
\ref{table_Ar}) and this difference is consistently observed along the
ablation cloud, as shown in the Figure \ref{Ablated_gas_T_compare}.
The reductions of the ablation rates for the neon and argon pellets by
atomic processes, $\sim15.6\%$ and $\sim25.3\%$ respectively, is
significantly larger compared to the case of deuterium pellets.  The
reason for this is well understood in terms of ionization energy
losses.

\bibliographystyle{unsrt}

\begin{thebibliography}{10}

\bibitem{Pegourie} B~Pégourié.  \newblock Review: Pellet injection
  experiments and modelling.  \newblock Plasma Phys. and Controlled
  Fusion, 49, R87, 2007.

\bibitem{Parks78} P.~B. Parks, R.~J. Turnbull.  \newblock Effect of
  transonic flow in the ablation cloud on the lifetime of a solid
  hydrogen pellet in a plasma.  \newblock {\em Phys. Fluids}, 21:1735,
  1978.

\bibitem{ParksSesBay00} P.~B. Parks, W.~D. Sessions, L.~R. Baylor.
  \newblock Radial displacement of pellet ablation material in
  tokamaks due to the grad-B effect.  \newblock {\em Phys. Plasmas},
  5:1968, 2000.

\bibitem{ICRU84} \newblock ICRU Report 37, Stopping powers and ranges
  for protons and alpha particles.  \newblock {\em International
    Commission of Radiation Units and Measurement}, 1984.

\bibitem{Ishizaki04} R.~Ishizaki, P.~B. Parks, N.~Nakajima,
  M.~Okamoto.  \newblock Two-dimensional simulation of pellet ablation
  with atomic processes.  \newblock {\em Phys. Plasmas}, 11:4064,
  2004.

\bibitem{SamLuParks07} R.~Samulyak, T.~Lu, and P.~Parks.  \newblock A
  magnetohydromagnetic simulation of pellet ablation in electrostatic
  approximation.  \newblock {\em Nuclear Fusion}, 47:103--118, 2007.

\bibitem{ParksLuSam09} P.~Parks, T.~Lu, and R.~Samulyak.  \newblock
  Charging and $E\times B$ rotation of ablation clouds surrounding
  refueling pellets in hot fusion plasmas.  \newblock {\em
    Phys. Plasmas}, 16:060705, 2009.

\bibitem{FT_code} R. Samulyak, J. Du, J. Glimm, Z. Xu, A numerical
  algorithm for MHD of free surface flows at low magnetic Reynolds
  numbers, J. Comp.  Phys., 226 (2007), 1532 - 1549.

\bibitem{RS7} S. Wang, R. Samulyak, T. Guo, An embedded boundary
  method for parabolic problems with interfaces and application to
  multi-material systems with phase transitions, Acta Mathematica
  Scientia, 30B (2010), No. 2, 499 - 521.

\bibitem{RS14} A. Hassanein at el., An R\&D program for targetry and
  capture at a neutrino factory and muon collider source, Nuclear
  Instruments and Methods in Physics Research Section A: Accelerators,
  Spectrometers, Detectors and Associated Equipment, 501 (1), 70 - 77.

\bibitem{RS15} R. Samulyak, Y. Prykarpatskyy, T. Lu, J. Glimm, Z. Xu, M.-N. Kim,
Comparison of heterogeneous and homogenized numerical models of cavitation,
International Journal for Multiscale Computational Engineering, 4 (2006),
No 3, 377 - 389.

\bibitem{TSTT_code} B. Fix, J. Glimm, X. Li, Y. Li, X. Liu, R. Samulyak, Z. Xu,
A TSTT integrated FronTier code and its applications in computational fluid physics,
Journal of Physics: Conference Series, 16 (2005), 471 - 475.

\bibitem{Zeldovich} Ya.~B. Zel$'$dovich and Yu.~P. Raizer.  \newblock
  {\em Physics of shock waves and high-temperature hydrodynamic
    phenomena}.  \newblock Dover, 2002.

\bibitem{KimSamZhang12} H.~Kim, R.~Samulyak, L.~Zhang, P.~Parks.
  \newblock Influence of atomic processes on the implosion of plasma
  liners.  \newblock {\em Phys. Plasmas}, 19:082711, 2012.

\bibitem{KimZhangSam13} H.~Kim, L.~Zhang, R.~Samulyak, P.~Parks.
  \newblock "On the Structure of Plasma Liners for Plasma Jet Induced
  Magnetoinertial Fusion,"
\newblock {\em Phys. Plasmas}, 20:022704 ,2013.

\end{thebibliography}

\end{document}